\documentclass[twoside,11pt]{article} 

\usepackage{graphicx}
\usepackage{epsf}
\usepackage{psfig}

 \newcommand{\noweb}{\mbox{ Noweb}}

 \newcommand{\prolog}{\mbox{\tt PROLOG}}

 \newcommand{\web}{\mbox{ Web}}
 \newcommand{\cweb}{\mbox{ Cweb}}
 \newcommand{\spider}{\mbox{ Spider}}

 \newcommand{\HP}{\mbox{HyperPro}}

 \newlength{\hhh}

\begin{document} 
\pagestyle{myheadings} 
\markboth{WLPE'01}{HyperPro}

\title{{\bf HyperPro\\An integrated documentation environment for CLP}\footnote{In A. Kusalik (ed), Proceedings of the Eleventh Workshop on Logic Programming Environments (WLPE'01), December 1, 2001, Paphos, Cyprus. COmputer Research Repository (http://www.acm.org/corr/), ** your cs.PL/0111046 or cs.SE/0111046**; whole proceedings: cs.PL/0111042.}}

\author{AbdelAli Ed-Dbali\\LIFO - Université d'Orléans\\BP 6759 - 45067
Orléans Cedex - France\\ {\tt AbdelAli.ED-DBALI@lifo.unv-orleans.fr}, {\tt
http://www.univ-orleans.fr/SCIENCES/LIFI/Membres/eddbali}\and
        Pierre Deransart\\INRIA Rocquencourt - BP 105 - 78153 Le Chesnay -
	France\\ {\tt Pierre.Deransart@inria.fr}, {\tt
	http://contraintes.inria.fr/\~deransar}\and
        Mariza A. S. Bigonha,
        Jos\'{e} de Siqueira,
        Roberto da S. Bigonha\\DCC - UFMG - Brésil\\{\tt \{mariza,jose,bigonha\}@dcc.ufmg.br}}

\date{}
\maketitle

\begin{abstract}
The purpose of this paper is to present some functionalities of the \HP\  System.
\HP\  is a hypertext tool which allows to develop {\em
Constraint Logic Programming } (CLP) together with their documentation. The
text editing part is not new and is based on the free software Thot.
A \HP\  program is a Thot
document written in a report style. The tool is designed for CLP but it can
be adapted to other programming paradigms as well. Thot offers navigation and
editing facilities and synchronized static document views.
\HP\ has new functionalities such as document exportations, dynamic views
(projections), indexes and version management.
Projection is a mechanism for extracting and exporting relevant pieces of
code program or of document according to specific criteria.
Indexes are useful to find the
references and occurrences of
a relation in a document, i.e., where its predicate definition is found and
where a relation is used in other programs or document versions and,
to translate hyper-texts links into paper references.
It still lack importation facilities.
\end{abstract}
 
%\keywords{CLP, literate programming, program version, structured editor,
%hypertext, documentation, specification.}


%\resume{Le but de cet article est de présenter certaines fonctionnalités du
%système \HP. \HP\ est un environnement de développement integré pour la PLC
%(programmation en logique et par contraintes) qui offre deux aspects du
%développement d'un programme\,: le codage en PLC et la documentation. Un
%document \HP\ est donc une combinaison de code et de sa documentation. La
%partie édition d'\HP\ est basé sur le système Thot. \HP\ a été conçu pour la
%PLC mais il peut aussi bien être adapté à d'autres paradigmes de
%programmation. \HP\ est un éditeur WYSIWYG qui offre des facilités de
%navigation hypertexte, de manipulation de versions d'un même programme,
%d'extraction de vues et projections selon des critères spécifiques. Il offre
%également l'exportation de document vers plusieurs formats courants (ps,
%html, latex, ...) ainsi que la gestion des index.}

%\motscles{PLC, programmation littéraire, gestion de versions, édition
%structurée, hypertexte, documentation, spécification}

\section{Introduction}
\HP\  is a documentation and development tool for Constraint Logic Programming (CLP) systems.
\HP\  helps to document CLP programs giving its users the possibility
to edit, in a homogeneous and integrated environment, a single program or different
versions of a program, comments about them, information for formal verification
and debugging purposes, as well as the possibility to execute, debug and test the
program or program versions as well. All the tests executed on a CLP program
and its history of development can therefore be integrated and consistently documented within an unique
environment.

\HP\ has two basic characteristics: version management and documentation. A particular aspect
of CLP is that the written programs are short and they can be tested on
multiple versions which differ only by small changes. In fact, for example,
the way constraints are ordered may drastically affect the efficiency and the
reachability of a solution. It is thus essential for the user to keep trace
of the many versions of a program (there is potentially an exponential number
of versions). In this approach of program development, documentation is at
least as useful as the code of the program
itself. It can be compared to elaboration of formal specification where documentation (or requirements) in
natural language plays a role at least as significant as the code.

The \HP\  system,
developed by the authors, make use of the Thot editor
and its API. 
Thot is a structured editing tool
with hypertext editing facilities. It was developed at INRIA and it is based
on the logical aspect of the document
(\cite{QV86}, \cite{QV92}, \cite{QV96}, \cite{QV95}).
Thot uses a meta-model that allows the description of several models of
documents, with their various presentations, through dedicated languages.
Thot is in fact very comparable with XML technology (\cite{xml}). Its language
$\cal S$ defining the structure of a document is similar to XML's DTD ({\em
Document Type Definition}). Its languages $\cal P$ and $\cal T$ (respectively
presentation and translation languages) could be compared with the XSL ({\em
eXtensible Stylesheet Language}) and XSLT (for {\em Transformation}).
We chose Thot because it provides an API which allows the easy creation of an
editor such as \HP\ and offers the possibility to integrate new applications
to existing ones. It has also a remarkable feature: the possibility to open
several views of the same document in synchronized windows.

This paper describes some of the \HP\ functionalities which are:
static synchronized views of the document; management of multiple versions of a program within the same document;
testing of different program versions;
test of these versions using external CLP systems; dynamic views (produced by
projections requests); indexes; exports of the document under different
formats for processing with other systems. This paper completes the
presentation of \HP\ published in \cite{DPal96} and presents the main novelties:
support for execution, exportations, indexes and projections. Projection is a mechanism for
extracting and exporting relevant pieces of code according to specific
criteria. Indexes are of three types: the traditional index of words, the
index of the relations cross references and the index of the program versions.
The two latter relate to the code parts and highlight the "two-dimensional"
hypertext structure of a document.

Thanks to the modularity of the implementation, with few changes, \HP\ can be
adapted to other languages such as the language C (cf. \cite{Mont98}).
 
In section \ref{sec_functionalities}, we survey basic \HP\ features whose
ideas where initially presented in \cite{DPal96}: static views, versions,
program execution and exportations.

Section \ref{sec_index} is dedicated to the indexes, section
\ref{sec_projections} to dynamic views (or projections) and we conclude this
paper by a comparison between \HP\ and other literate programming systems.

%A { predicate definition} contains : {\em informal
%comments}, {\em assertions} and {\em packet of clauses}, where,
%{ informal comments} are a sequence of paragraphs and
%{ assertions} are a sequence of lines of text and are optional.
%{ Packets of clauses} can be Prolog or {\tt clp(FD)} clauses. Clauses
%include directives, goals, facts and rules. The predications in the clauses
%body may have references or not to the relations which define them. This will
%actually define the current version of a program in a \HP\  document,
%as explained below.
%Every relation definition must have at least one predicate definition.
%When a predicate definition is introduced, the fields for informal comments
%and packet of clauses are initially empty. The optional fields are presented
%if the user explicitly indicates so. However, the obligatory fields do not
%need to be filled in immediately after their insertion in the document.
 
\section{HyperPro Functionalities}
 \label{sec_functionalities}
 
This section introduces the most important functionalities of \HP,
which are the static views of different
parts of the document, program versions, their testing and document export
in various formats.
 
The \HP\ document looks like a report. It must have a
title, a sequence of at least one section, a table of contents and a words
 index. Optionally, it can contain the date,
the authors' names and their affiliations, keywords, bibliographical
references, annexes and other indexes.
\HP\  defines, also, a special style for logic programming, so
 a paragraph, may be also a {\em relation definition}. At least one
relation definition must be present in a \HP\  document.
A relation definition contains a {\em relation title} and a list of
at least one {\em predicate definition}. The relation title is defined by
a predicate indicator, that is the predicate name and its arity, or just a name,
since a packet of goals or directives can be seen as a special case of
relations. The relation title contains also a reference to a predicate
definition, followed optionally by a list of references to predicate
definitions that define a unique program or different {\em versions} of
the program.
 
\subsection{Static views}

Thot allows the user to define views of his document, such that,
chosen portions of it are presented separately in a view. Views are
synchronized in such a way to facilitate selecting, moving and editing
consistently and easily in a particular view as much as in the main view.
\HP\  provides several different views of the same document:
 the main view that shows the whole document, the table of contents,
 the informal comments on relations,
 the assertions on relations,
 the relations code which are a collection of packets of clauses corresponding
 to different versions of the same relation.
The main view and the table of contents are displayed automatically, when a
document is open. The other views are
shown on demand by the user through the appropriate menu bar.
 
By clicking somewhere in the main view, like in one of the other opened views,
all the views are synchronized simultaneously to present the corresponding
part of the referred text.

\subsection{Program Versions}
 
The possibility of being able to manage various versions of a program during
the development and documentation is a significant feature offered by \HP. This
possibility finds all its interest in the case of the Constraint Logic
Programming. Indeed, it is interesting to be able to integrate in the same
document several versions which, for example, differ only by one clause or are
completely different because they are based on different algorithms.

%A { program} is a set of packets of clauses. A document may
%contain only one program. It is up to the user to decide how the
%document will be organized, defining his program throughout sections and subsections. The user can then define a program by selecting
%conveniently predicate definitions in the document, and putting references on
%them. \HP\   provides
%the means to
%define, document and correctly manage his program, but it does not give means
%to solve any conflicts that could appear during these processes. Indeed, since
%two logic relations can be defined with the same name and the same arity anywhere in
%a document, this should be avoided by the user within the same program.
%\HP\  allows the user the possibility to manually define and
%automatically test, view and  export his program.
 
%The possibility of having different versions of a program
% is another important feature when defining, documenting and
%developing programs. Specially when developing and documenting CLP programs,
%the user needs to define, test and document different {\em versions}
%of his program. For that, he may define for any relation, different
%versions of its predicate definitions which are documented and managed with
%the same utilities used to define the program. Therefore, a program may have
%several different versions which can be defined, named, viewed and tested as
%much as any program. Indeed, { program versions} are programs which differ
%at least in one clause.
% 
%\subsubsection{Defining Programs and their Versions}
%\label{311}

Every relation is defined by at least one packet of clauses. Among these packets we distinguish a particular one which is the current packet defining the relation. It is called the {\em current predicate definition}
({\em c.p.d.}). The c.p.d. is pointed by a hypertext link placed in the
relation title. This link is called the {\em current predicate reference} ({\em[c.p.r.]}). The [c.p.r.] can be changed to point to any other predicate
definition the user wishes, within the same relation definition. The
[c.p.r.] are
obligatory since they define the current or the default predicate definition
for each relation in the document.
\begin{figure}[htb]
\centering
%\leavevmode
%\epsfbox{versao2CLEI.eps}
\includegraphics[height=11cm,width=12cm]{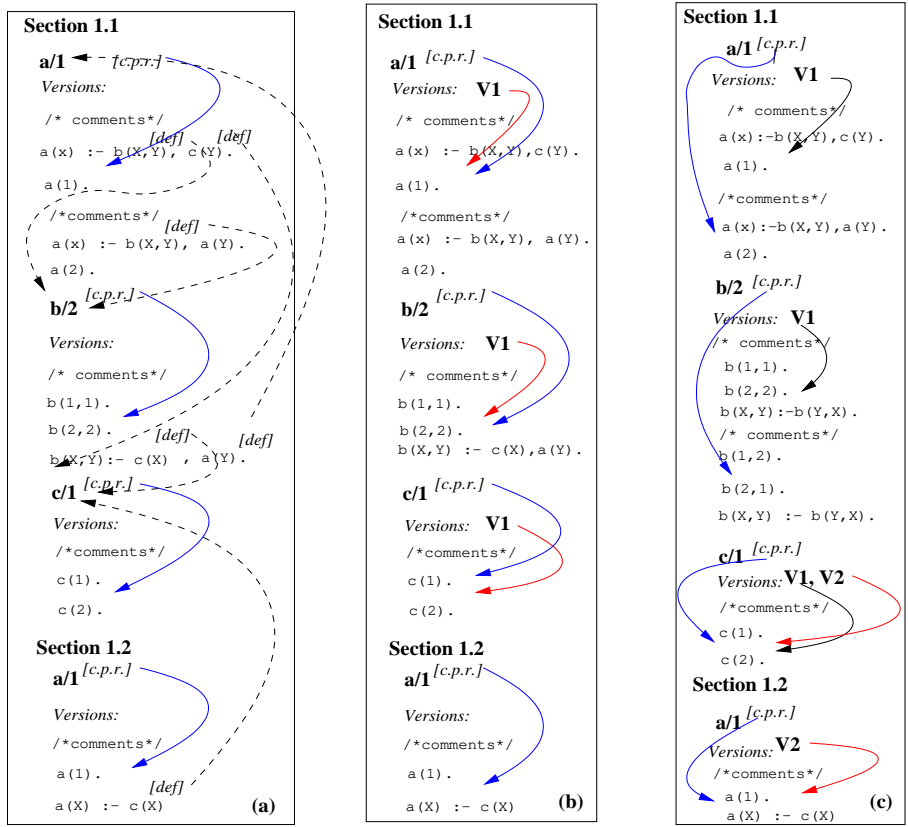}
\caption{(a) Some Relations, [c.p.r.] and Version References (b) After Naming Version {\tt V1} (c) After naming Version {\tt V2}}
\label{fig2}
\end{figure}
 
A program is defined when it has appropriate references to the relations which
compose it.
 Once all the relations and their c.p.d. for a program are given, the user
defines the program by putting a hypertext reference to the relation which
defines the predication in every predication in the body of all relation's
c.p.d. These references are called {\em definition references} ([def.]). The user
does not have to put [def.] to predications in the body of the c.p.d. which are
 direct recursive calls.
 
Figure \ref{fig2}(a) shows some relations defined for a program, with the
[c.p.r.] presented in the relation titles composing the program and
the [def.] in the predications at the body of the c.p.d. The
document
has two sections, section {\tt 1.1}, which defines
relations {\tt a/1}, {\tt b/2}, and {\tt c/1}, and section {\tt 1.2}, which defines relation {\tt a/1}.
Only relation {\tt a/1} in section {\tt 1.1} has
two predicate definitions defined for it. The [c.p.r.] for relation {\tt a/1}
points to its first predicate definition. The [c.p.r.] are represented by full lines
in the figure. The dashed lines represent the definition references, which
are presented in the text of the document as {[def.]}. As it is shown, they
point to the relation titles they refer to. Note that there are no [def.] for
predications recursively referring to the relation they are defining.
 
The [def.], together with the version naming utility, allows the user to
effectively manage and document different versions of the program
 in the same document which can
have different relations with the same name and arity. Conflicts between
relations are therefore solved and managed explicitly and manually by the user
with [def.]. Different versions of the program
 defined in a document are distinguished from
each other by their names and are managed with the corresponding naming reference.
The naming reference
allows the user and the utilities based on it to follow and retrieve all
predicate clauses of every relation in a program as defined by the chain of
[def.] of every relation linked in it.
 
To name a version, once the [def.] for a program are inserted
for the predications in the body of the c.p.d. as explained above, the user
selects the  utility {\em name a version} in the {\em Tools} menu.
A window opens and the user enters a name for the version. The user
then chooses the first
%\footnote{This constraint in using this utility made easier its implementation.}
relation composing his program and a named reference is put then automatically in the version bar, after
the title of every relation
composing the version, pointing to the respective c.p.d. of each relation
in the program.
 
Figure~\ref{fig2}(c) shows two different versions named { V1} and { V2}
defined in a document which has, among others, two different relations with the
same name and the same arity. This figure shows the final state after defining
both versions.
However, the user first named version V1, as presented in Figure
\ref{fig2}(b) by choosing the  utility
{\em ``name a  version''}  in the {\em tools} menu, giving the
version the
name V1, and finally choosing relation {\tt a/1}, which is the first relation in
his version. The pointers for { V1} point to the c.p.d., i.e. to the
same predicate definitions pointed to by the [c.p.r.].
 
Afterwards, the user entered a new section, numbered 1.2, where he
defined relation {\tt a/1}. In its predicate definition, the second clause for
{\tt a/1} refers to relation {\tt c/1}, previously defined in section {\tt 1.1}.
Therefore, he puts a [def.] after the predication {\tt c(X)}, which points to
relation {\tt c/1} in section {\tt 1.1}. Then the user proceeds to name this new
version, choosing for it the name of V2. After choosing relation {\tt a/1} in
section {\tt 1.2}, the naming reference { V2} appears in the version bar after
its title, which points to the
c.p.d. pointed to by the [c.p.r.], as it appears in the title of relation {\tt c/1}
as well, since it also became part of version { V2}.
 
The user is able to delete a version by simply indicating its name. The system then removes automatically all the pointers defining it as well as the name of this one accompanying the relation titles in which it appears. 
 
%Whenever a program is defined and the user changes the [c.p.r.] of a relation
%composing it to point to another predicate definition, the user is defining
%a new program version. This is the simplest way to define a program version.
%However, the definition is not entirely completed, since the user has to
%insert [def.] in the predications at the body of the c.p.d. and then name
%it. This is done with the naming utility, as explained above.
%Note that,
%although the [c.p.r.] changed for the relation the user modified, all the naming
%references
%which pointed to the previous c.p.d. still define the corresponding versions.
%It is only a new  version that is being defined and named, and all
%previous versions depending on other predicate definitions for this relation
%are still properly defined.
 
In Figure \ref{fig2}(c), we still consider the same  \HP\ 
document presented in Figure \ref{fig2}(b). However, here the user modified
the [c.p.r.] for relation {\tt a/1} in section {\tt 1.1} and added a new predicate
definition for relation {\tt b/2}, modifying also its [c.p.r.] so as to point
to its second predicate definition. However, the naming reference { V1}
previously defined still points to the previous predicate definition.
The [def.] are not shown in this figure so as not to complicate the figure
excessively. 
Then, the user proceeds to define this version with the utility {\em name a version},
giving it the name { V1} (or other name) and choosing the relation {\tt a/1}.
Then, a naming reference { V1} is automatically put in the version bar after
every title of all the
relations of the version, along the [c.p.r.]/[def.] chain.

\subsection{Program Testing}
 
Once a version is defined, the user can test it, take a view of
it, verify it syntactically or export it.
To test a program or a version the user should specify the interpreter for
a language he wants
to test, from the available ones, in a menu. Then, he
can choose one of the following test modes:
 program/version mode;
clause/relation mode;
automated recursive mode.
 
In the {\tt program/version} mode, the user selects a named version wherever it appears
in the document and then \HP\  opens a test window where the
previously chosen interpreter is run and where the version is loaded, starting
from the beginning of the program.
The {\tt clause/relation} mode allows the user to select any packet of clauses
or relation which are loaded in the test window. If a relation is selected,
it is loaded from its c.p.d.
In the {\tt automated recursive} mode, the user selects one relation in the document and the test utility loads
it, following the [c.p.r]/[def.] chain in which the selected predicate definitions
are included, starting from the the selected relation. It is the simplest
way to test a small program or any set of few predicates, without having to
explicitly define a program or a version.
 
The user can then run and test his program, version or packet of clauses in the
test window. Therefore, the only thing that changes when testing a program in
each mode is what is loaded from the document in the test window.
%The user can also import the program results into the document.

%\subsection{Syntactic Verification}
% 
%\HP\  presents the user the possibility to syntactically verify
%whether any version of his program or predicate definitions is correct or not,
%according to some pre-defined syntax, chosen from a menu. \HP\ 
%offers three ways to do syntactical verification, in the same
%way it is done for program testing, according to the selection
%mode:
%program/version mode;
%clause/relation mode;
%automated recursive mode.
% 
%In the
%program/version mode, a program or version is selected by its naming reference or name;
%in the clause/relation mode, a predicate definition is selected, either
%directly or by selecting a relation, and in this case, it is the c.p.d. for the
%relation which is selected; and in the automated recursive mode, a relation is
%selected and all
%the predicate definitions in its [c.p.r.]/[def.] chain are selected.  In all modes,
%the selected clauses are verified by a syntactical verifier for the language
%chosen and its results are displayed in a separate window.
%The choices for syntactical verifiers include Standard Prolog (\cite{DEC96}) and
%{\tt clp(FD)} in \HP, and a future version should include
%syntactical verifiers for Prolog IV and Chip.
 
\subsection{Document Exporting Facilities}
 
\HP\  allows the user to export the whole document in four different
formats:
 ASCII,
 \LaTeX, Postscript and
 HTML.
The ASCII exporting facility simply makes a dump of ASCII codes of the document
into a file. It is useful during \HP\  development especially when the
structure of the \HP\ document is changed.
The \LaTeX exporting facility dumps a \HP\  document into a file in
atex format. The logical structure of the original document is entirely
reflected in the \LaTeX file, except for the hypertext links, for obvious
reasons. However, the indexes are mirrored in the \LaTeX document, so that the
hypertext version of the original document is faithfully rendered in paper,
as much as possible. The table of contents is maintained in the \LaTeX version
of the document.
The HTML exporting facility makes the original \HP\  document
 viewable by any available web browser, where the original
hypertext links appear as such in the HTML version of the document.
It is also possible to export only part of the document like a version or a packet of clauses using appropriate menus.

\section{Indexes}
\label{sec_index}
When presenting any hypertext document in two dimensions, as on paper, we
obviously loose the hypertext links together with the linked information, and
the hypertext facilities as well. The only way to render these informations again in
two dimensions is by means of indexes. But this is not the only reason to build
indexes. In fact, an index can present the information of all linked
information linearly, allowing the user to access any node of the linked
hypertext structure instantly, and not only through the linear link sequencing.
Another advantage of building indexes, since the \HP\ editor allows the
presentation of any part of a document in a separate view, is that we can present,
through the index-based projections in \HP, parts of the original
document by selecting the appropriate information directly from the indexes in
a \HP\  document. We present each of the \HP\  indexes in the next
subsection and the index-based projections in the following one.
 
A \HP\  document has two hypertext structures:  one linking together
information about relations and the other connecting
relations together in program versions.%, as explained in Section \ref{311}.
There is an index for
each of these hypertext structure, called {\em Cross Reference Index} and
{\em Versions Index}, respectively. Each of these indexes is built
independently through the Tools menu presented in the \HP\ editor. Once an index
is built, the \HP\  system opens a new view for the corresponding index.
 
\subsection{Cross Reference Index}
\begin{figure}[htb]
\centering
\includegraphics[height=11cm,width=12cm]{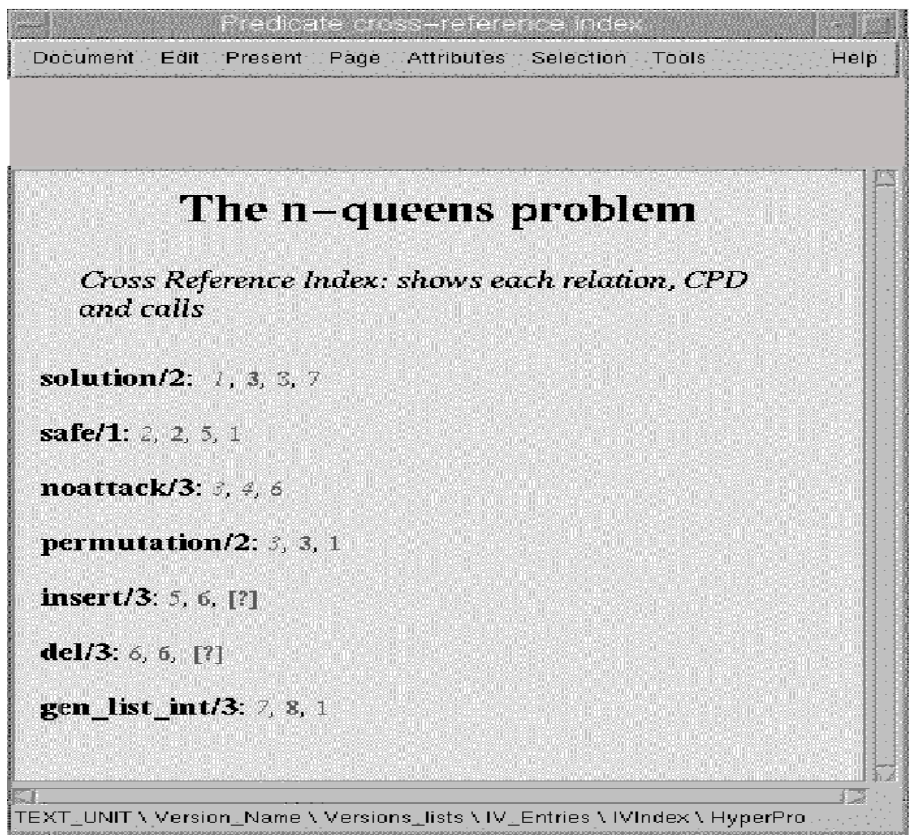}
\caption{\label{cri} Cross reference index view image}
\end{figure}

The Cross Reference Index indicates the page number of every relation appearing
in the document, the page number where the relation's current predicate
definition is found, and the page numbers where the relation is used in other parts of
the relations defined in the document, independently of versions. It is an
absolute index of relation definitions. Each of these page entries is
presented differently in the \HP\  document, as well as printed: the
relation position in the document is shown in italic, the current predicate
definition in bold, and each entry where the relations used in normal font.
Figure \ref{cri} shows a Cross Reference Index built for a \HP\  document.
 
Since the index page entries are hypertext links, the user can access the
corresponding part of the original document by just clicking its index page
entry, and the main document view will present synchronously the part of the
document corresponding to the entry.

\subsection{Versions Index}
\begin{figure}[htb]
\centering
\includegraphics[height=11cm,width=12cm]{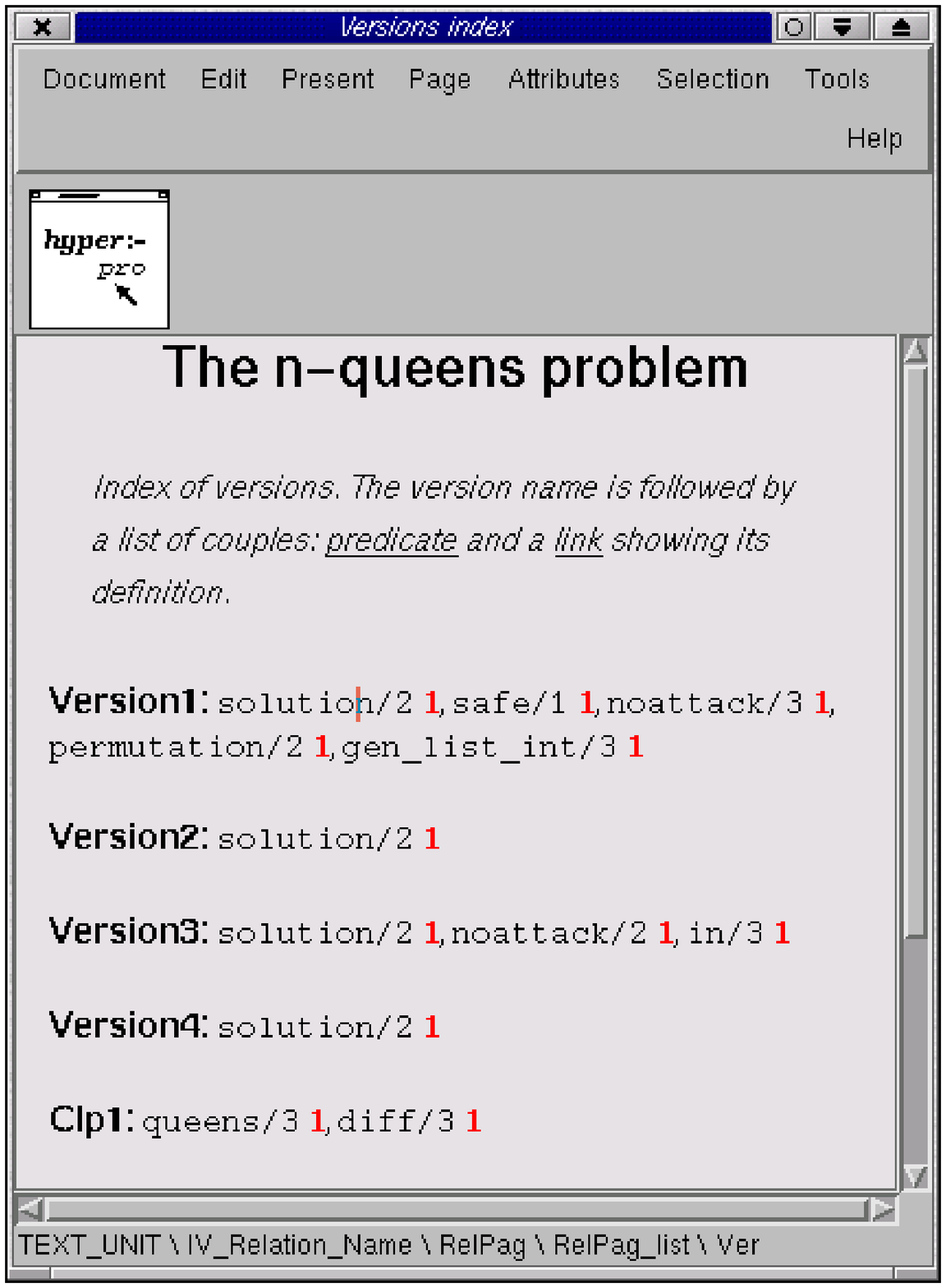}
\caption{\label{ver} Version Index View Image}
\end{figure}

The Versions Index presents, for each version, all the relations which are part
of it, together with the document page number where the relation is defined.
Figure \ref{ver} shows a Versions Index.
 The first entry indicates where the version was firstly
defined, i.e. in which page of the document is found the predicate definition
pointed to by the first named references appearing in the document, as all corresponding
predicate definitions pointed to by the named references for each relation included in the
version as well.

As for the Cross Reference Index, each page number of a relation
entry in the Versions Index is a hypertext link to the corresponding place
where the relation is defined. Even if there are more than one relation
with the same name and arity, both defined in the same document page, each
participating of a different version, the main view will show only the
corresponding entry for each relation page entry link clicked.

\section{Dynamic views}
\label{sec_projections}
 
\HP\  allows the possibility to project and view
separately some parts of the document selected by the user following some
criteria.
 
The dynamic views (called also {\em projection views})  \HP\  offers are:
 manual projection view;
 automated projection view;
 version view;
 recursive projection view;
index-based projection view.

A { projection  view} displays selected portions of the document in a
separate view. The selection processes depends on the projection wanted.
Projections differ from static views on the selection process, which are already
incorporated in Thot's machinery in the case of static views, but have to be provided
dynamically by the \HP\  implementors in case of projections.
 
In the {\em Manual Projection View Mode} the user is free to select any part of
the document, called elements, to be viewed.
The granularity of the selected element is defined to be
paragraphs and relation definitions. Therefore, when the user clicks in any
of these
elements or in their descendants, the projection will show the whole paragraph or relation definition.

In the {\em Recursive Projection View} the user selects one or more relations
and
\HP\  shows in a separate view all the packets of clauses in the
[c.p.r.]/[def.] chain in which the relation is inserted, including obviously the
packet pointed to by its own [c.p.r.]. This projection view is rather useful to
help the user to detect which predications have not yet its [def.] set, or to
find out where it is set to.
 
In the {\em Automated Projection View Mode}, the user gives some regular
expression and \HP\  shows in a separate view all the portions of the
document where the regular expression appears.
The smallest granularity for this projection is a word.
 
The {\em  Version Projection View} shows the user, in a separate view, all the
packets of clauses composing a program or version chosen by the user in the
document.
 
\begin{figure}[htb]
\centering
\includegraphics[height=14cm,width=12cm]{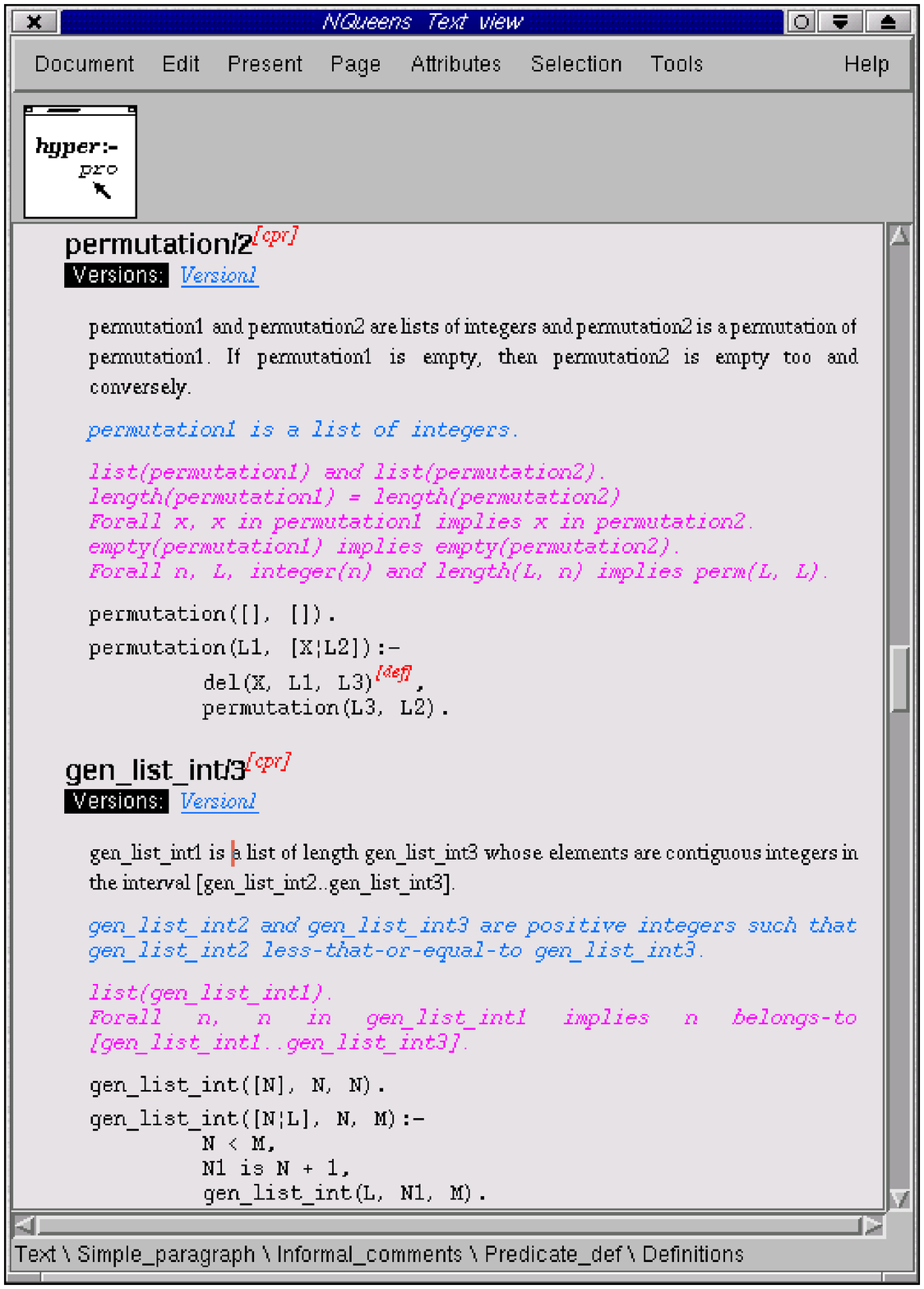}
\caption{\label{pvi} Version Index Projection View Image}
\end{figure}

{\em Index-Based Projection View:}
 \HP\  offers
a projection associated to indexes. Once an index view is built, as explained
in Section \ref{sec_index}, the user can choose in the Tools menu the {\tt Projections}
entry which will
show  a sub-menu where the different projections are presented. If the
user chooses the {\tt Index based} entry, clicking on an index
entry, either {\tt Versions} or {\tt Cross Reference}, \HP\  will automatically show in
another separate view what is related to the entry chosen in the index view.
For instance, if the user clicks on a {\tt Version Index view} entry, the projection
will present in its view all the relations which are part of that version
entry, and only them. Clicking on any part of the projection view will
synchronously present the corresponding part of the document on the main view.
In the case of the Cross Reference Index, the projection will show the relation
corresponding to the entry the user clicked on. Figure \ref{pvi} presents the
image of a projection view for an entry of the Version Index shown in Figure
\ref{ver}.

\section{HyperPro Interfaces}
 
In this section we present the user's schematic interfaces such as menus and
sub-menus for the functionalities and utilities that should be present in
\HP. Some functionalities and utilities appear in specific Thot
menus, like the export, views and index facilities. However, the manual and
automated projection view, the version or program view, the recursive view and
the index-based projection views facilities appear in the Tools menu, as
the others utilities as well. The reason is that Thot allows anyone to include
his own facilities using the Thot toolbox.
%which are accessible through the Tools menu.
 
Views are selected in the menu {\em Views}. There the user chooses the
entry point to the sub-menu {\em Open a view...} and in there, he chooses the
view he wants to open, including the indexes views. The views the user can open
are, therefore:
 the main view of the whole document;
the table of contents;
 view of the comments;
 view of the assertions;
 view of the clauses;
 view of indexes.
 
The exporting facilities are accessed by the {\em Document} menu, and the
{\em Save as...} sub-menu. A window opens where the user can then choose the
exporting facility he wants to use.
We will not present here the index facility interface.
%The reader can find about it and how to use this facility in \cite{jose,jose1}. 
%For projection consult
%\cite{BMal99} to see for each projection its dialog box.
Tools menu offers the user access to the following utilities:
Make indexes,
Versions,
Projections,
Tests,
Syntax verification and
\HP\  preferences.
 
All facilities interface is done through {\em communications windows}, where
the user controls the utility and where the data input is done, and some
utilities may open a specific window to work as their output interface, as explained
above.

\section{HyperPro Versus Other Systems for Documentation}
\label{statearta}
 
%Logic program development has been considered by different authors
%from different point of views. Most of them consider program
%development as a transformation process from a specification to an
%efficient Prolog program. Deville \cite{DY90} starts from first order
%logical formulas and ``mode'' declarations. More generally a mode
%declarations can be viewed as type declarations.
%Most of the systems intend to help the programmer in developing
%correct programs, or verifying afterwards that the program satisfies
%some properties. In logic programming different kind of proof systems
%have been designed. In \cite{LD91} one of these system is described.
%In \cite{DM93} a systematic
%approach of logic program validation is presented. Some of the ideas
%have been implemented in the system LDS2 described in \cite{RD94}
%and used to define a methodology for writing specifications in logic
%programming style \cite{ED92}.

At the current state-of-the-art, there are no satisfactory tools or
widely accepted methodologies for documenting \prolog\ programs.
Donald Knuth introduced literate programming in the form of \web, his
tool for writing literate Pascal and C programs   \cite{knu83a,knu92a}.
The  philosophy
for documenting
Pascal  or {\bf C} programs
apparently offers
the basis to establishing a methodology to document programs in
the logic programming paradigm, but it seems not sufficient as we will see.
 
In the context
of \web-like literate programming systems developed since 1984,
the following are the most important documentation
systems:
1) Knuth's \web\ for Pascal and {\bf C} \cite{knu83a};
2) Ramsey's \noweb\ \cite{Ram94a};
3) Thimbleby's \cweb\ \cite{Thim86}, a variant of Knuth's \web;
4) Ramsey's \spider\ \cite{Ram89} which is a \web\ generator.
The basic idea behind {\em Literate Programming}
is that programmers should use three  languages:
a typesetting language, such as \LaTeX\ ; a programming language,
such as Pascal, and a language which allows
flexible combination of the typesetting and the programming texts into
a single document.
Thus, a literate program contains pieces of programs  interleaved
with
descriptive texts.
A literate programming system integrates these languages by
providing
tools to extract and process, from the input files, program texts
and to generate documents containing summaries, index tables, cross-references, etc.
 
\cweb\ \cite{Thim86} is a tool to produce program
documentation in
a combination of C, the programming language, and {\em troff},
a text-formating language.
The combined code and documentation can be processed and
possibly typeset to result in a high-quality presentation
including a table of contents, index,
cross-referencing information, and related typographical conventions.
\cweb\ differs from  \web\  mainly in the choice of
languages: \web\ is based on
Pascal and \TeX\ instead of C and {\em troff} or {\em nroff}.
 
\spider \cite{Ram89}  was designed for developing verified Ada
programs.
The difficulty of using \web\ directly is that the intended target
programming language is SSL
(language for specifying structured editors),
 and the
only languages for which \web\ implementations were available were
Pascal and C.
\spider\ is a \web\ generator, akin to parser generators.
Using \spider\ the user can build a \web\ without understanding the details of
web's implementation, and
can easily adjust that \web\ to change as a language definition changes.

Norman Ramsey has proposed a new literate programming system,
called \noweb\ \cite{Ram94a}, which is intended to be a simple and extensible tool.
It was developed on Unix and can be ported to non-Unix systems provided
that they can
simulate pipelines and support both ANSI-C and either awk or icon.
\noweb\ can also work with HTML, the hypertext markup
language for Netscape and the World-Wide Web.
A \noweb\ file is a sequence of {\em chunks}. A chunk may contain code
or documentation and may appear in any order.
Code chunks contain program source code and references
to other code chunks. Several code chunks may have the same name.
Ramsey claims that \noweb\ is simpler than Knuth's \web\ due to its
independence of the target programming language, but it also means that \noweb\
can do less.
At last, \noweb\ works with both plain \TeX\ and \LaTeX.
The system is extensible in the sense that new tools can be easily added
to it, requiring no reprogramming.
Its Weave tool preserves white spaces and program indentation when
expanding chunks. Theses features are necessary to document program in languages
like Miranda and Haskell, in which indentation is significant.

An other documentation generator called {\em lpdoc} was developed by Manuel
Hermenegildo team \cite{lpdoc-cl2000}. It is designed for (C)LP and more
specifically Ciao programs.
lpdoc generates a documentation, in
various formats, automatically from one or more source files for a logic
program. The quality of the documentation generated is enhanced when using the
Ciao assertion language and the machine-readable comments (using the
'comment' directives). 

%To conclude, no present system has gone beyond the experimental
%stage or beyond the capacity to handle small programs. Programming in the
%large with such systems is an objective still far to be
%reached. The problem of documenting while developing the programs is
%marginally considered.

All the systems described above can not be considered as integrated environments for
development and documentation. They do not offer a WYSIWYG tool to handle
all literate programming aspects in an uniform way.

Our purpose while developing \HP\ was to offer a tool which permits to record all the
experiences accumulated when developing an application based on
constraint logic programming, and maintaining it: programming, documenting,
debugging, performing verifications, testing.
The high level of expressiveness of constraint logic programming
makes possible to consider a program as an executable
specification.
\HP\  presented in this paper consider that a CLP program is a
document written with a methodology which takes into
account the peculiar aspects of logic programming on one side. On the
other side, it has the flexibility of a textual document.
All the information concerning the program development and its
maintainance is recorded in this document.

\section{Conclusion}
 
We have presented a system based on the hypertext system
Thot \cite{QV95}, called \HP, whose purpose is to facilitate logic programs development. It
offers several facilities to view and handle documents at different levels
of abstraction and from different point of view. Particularly,
\HP\  aims to document CLP programs giving its users the possibility
to edit, in a homogeneous and integrated environment, different
versions of programs, comments about them, information for
formal verification
and debugging purposes, as well as the possibility to execute, debug and test the
programs as well. All the attempts and development history of CLP programs
can therefore be integrated and consistently documented within a unique
environment gathering together a hypertext editor, different CLP interpreters
and syntactical verifiers, as different debugging and verification tools as
well.
It also possesses
several functions for exporting the document in different formats such as:
html, latex, ascii, and producing
projections, which are especial excerpts of the document, according to different
criteria, such as, the program goal, pieces of code directly marked by the user,
program versions, etc. \HP\ is convenient to create and maintain new programs.
However, it still need program importation facilities in order to be able to
incorporate and handle already existing programs.

%\bibliographystyle{plain} 
%\bibliography{hyperpro}

\begin{thebibliography}{10}

\bibitem{DPal96}
P.~Deransart, R.~Bigonha, P.~Parot, M.~Bigonha, and J.~de~Siqueira.
\newblock A hypertext based environment to write literate logic programs.
\newblock In {\em I SBLP}, pages 1--16, 1996.

\bibitem{lpdoc-cl2000}
M.~Hermenegildo.
\newblock {A} {D}ocumentation {G}enerator for {(C)LP} {S}ystems.
\newblock In {\em International Conference on Computational Logic, CL2000},
  number 1861 in LNAI, pages 1345--1361. Springer-Verlag, July 2000.

\bibitem{knu83a}
Donald Knuth.
\newblock The \texttt{web} system of structured documentation.
\newblock Technical Report 980, Stanford Computer Science, 9 1983.

\bibitem{knu92a}
Donald Knuth.
\newblock {\em Literate Programming}.
\newblock Number~27 in CSLI lecture notes. Center for the Study of Language and
  Information, 1992.
\newblock pages 349--358.

\bibitem{Mont98}
S.~Montagne.
\newblock Hyperpro(c) : un environnement pour la programmation en c.
\newblock Rapport de stage de fin d'études, 1998.

\bibitem{QV95}
V.~Quint.
\newblock The thot user manual.
\newblock Technical report, INRIA, 1995.

\bibitem{QV96}
V.~Quint.
\newblock The languages of thot.
\newblock Technical report, INRIA, 6 1996.
\newblock translated by Ethan Munson.

\bibitem{QV86}
V.~Quint and I.~Vatton.
\newblock {G}rif\,: an interactive system for structured document manipulation.
\newblock In {\em International Conference on Text Processing and document
  Manipulation}, pages 200--213. Cambridge University Press, 11 1986.

\bibitem{QV92}
V.~Quint and I.~Vatton.
\newblock Hypertext aspects of the grif structured editor: Design and
  applications.
\newblock Technical report, INRIA, 7 1992.

\bibitem{Ram89}
Norman Ramsey.
\newblock Literate programming: Weaving a language-independent \texttt{web}.
\newblock {\em Communications of the ACM}, 32(9):1051--1055, 9 1989.

\bibitem{Ram94a}
Norman Ramsey.
\newblock {\em The noweb Hacker's Guide}.
\newblock Princeton University, 9 1992.
\newblock Revised 08/1994.

\bibitem{Thim86}
H.~Thimbleby.
\newblock Experiences of `literate programming' using \cweb\ (a variant of
  knuth's \texttt{web}).
\newblock {\em The Computer Journal}, 29(3):201--211, 1986.

\bibitem{xml}
W3C.
\newblock Xml w3c recommendation.
\newblock http://www.w3c.org/XML.

\end{thebibliography}

\end{document}